# Spatiotemporal Deep Learning Network for Photon-Level Block Compressed Sensing Imaging

**Changzhi Yu[1, 2], Shuangping Han[1, 2, *], Kai Song[1, 2], Liantuan Xiao[1, 2, 3, *]**

[1.] College of Physics and Optoelectronics Engineering, Taiyuan University of Technology, Taiyuan, Shanxi 030024, China;

[2.] Shanxi Key Laboratory of Precision Measurement Physics, Taiyuan University of Technology, Taiyuan, Shanxi 030024, China;

[3.] State Key Laboratory of Quantum Optics Technologies and Devices, Institute of Laser Spectroscopy, Shanxi University, Taiyuan, Shanxi 030006, China.

* hanshuangping@tyut.edu.cn　(S. H.); xlt@sxu.edu.cn (L. X.)

## Abstract

In this paper, we propose a spatiotemporal deep learning network for photon-level Block Compressed Sensing Imaging, aimed to address challenges such as signal loss, artifacts, and noise interference in large-pixel dynamic imaging and tracking at the photon level. This approach combines information in the time and frequency domains with a U-Net-LSTM deep learning model, significantly improving the restoration quality of dynamic target images at high frame rates. The experimental results demonstrate that dynamic target imaging with an average photon number of less than 10 per pixel can be achieved using 16-channel parallel detection, where the pixel size is $256\times256$ and the frame rate is 200 fps.. Compared to conventional Block Compressed Sensing Imaging, this method increases the peak signal-to-noise ratio to 38.66 dB and improves the structural similarity index to 0.96. In the presence of a dynamic scattering medium and a static complex background, we successfully achieved imaging and tracking of two targets undergoing complex motion. Even in scenarios where the targets overlap or obstruct each other, we can still reconstruct clear images of each individual target separately.. This method provides an effective solution for large-pixel dynamic target recognition, tracking, and real-time imaging in complex environments, offering promising applications in remote sensing, military reconnaissance, and beyond.

**Keyword**：Spatiotemporal Deep Learning, Block Compressed Sensing Imaging, Deep Learning, Multi-Dynamic Targets

## 1. Introduction

With the ongoing development of fields such as satellite remote sensing, astronomical observation, and non-destructive testing, the demand for high-quality imaging under weak-light conditions has significantly increased[1-7]. Photon-level single-pixel imaging, utilizing single-photon detectors, enables the reconstruction of high-resolution images even in challenging weak-light environments[8-11]. However, existing technologies face the challenge of slow imaging speeds due to sampling time constraints, which limits their practical application in rapidly changing scenarios. Therefore, improving imaging speed and dynamic response capability has become a critical focus in the advancement of these technologies. Block Compressive Sensing (BCS) is an approach that enables efficient signal acquisition and processing by dividing the image into small blocks and performing Single-pixel Compressed Sensing (SCS) imaging in parallel[12, 13]. In contrast to traditional global reconstruction techniques, the block-based strategy not only simplifies the computational process but also accelerates imaging speed significantly through parallel processing[14, 15]. BCSI offers significant advantages in enhancing imaging speed and reducing computational complexity, making it especially suitable for dynamic monitoring that demands quick responsiveness[16, 17]. However, this approach still encounters several challenges. Firstly, some information may be lost because of the sparsity of sampling[18]. Secondly, due to limitations in fabrication techniques, unused gap regions exist between the detector's blocks, particularly in low-pixel detectors, resulting in low information utilization and the introduction of blocky artifacts[19]. These artifacts have a more substantial impact on image quality and reconstruction accuracy, particularly when there are significant differences between blocks. Consequently, minimizing the effects of signal loss and artifacts while enhancing imaging speed remains a critical challenge in the BCSI domain[20].

In recent years, deep learning (DL) has been widely applied to address complex challenges across various fields, particularly in image reconstruction and compressive sensing, where it has demonstrated significant advantages.. In BCSI, DL effectively compensates for information loss due to undersampling, successfully recovering high-quality images, especially at low sampling rates. By combining feature learning with reconstruction, these methods optimize both image quality and computational efficiency, particularly in high-dimensional compressive sensing problems[21, 22]. In image recovery, DL provides solutions for optimizing acquisition strategies and enabling non-iterative reconstruction. Adaptive strategies incorporating prior image knowledge reduce computational load, while multi-channel sampling combined with deep learning enhances both speed and accuracy in dynamic imaging systems[23, 24]. Moreover, multi-channel sampling with DL improves both speed and accuracy in dynamic imaging systems. Integrating array sampling with DL allows high-quality recovery under low sampling rates, boosting performance in dynamic scenes[25]. In dynamic scattering medium imaging, DL enables real-time, non-invasive recovery through turbulent water and natural haze, overcoming the limitations of traditional methods[26]. Despite these advantages, DL in image reconstruction still faces challenges. Most research focuses on single-frame recovery, but poor-quality or information-deficient images limit effectiveness. In multi-target scenarios, occlusions and overlaps hinder the simultaneous recovery of detailed target information, leading to interference and degraded reconstruction quality.

In this paper, we propose a spatiotemporal deep learning model for BCSI-based dynamic imaging to improve image quality and frame rate. This scheme effectively exploits the differences between frames, thereby eliminating different types of noise during the imaging process. Simultaneously, it addresses the signal loss caused by block-based and sparse sampling, enabling the comprehensive acquisition of dynamic target information. Furthermore, the U-Net-LSTM DL model is further employed to further restore the images. Experimental results demonstrate that, in low-light environments with an average photon count per pixel less than 10, the 16-channel parallel detection used in this approach enables dynamic target detection at 200 fps for 256×256 pixel images. The PSNR increases from 6.14 dB (in traditional BCSI) to 38.66 dB, significantly improving image quality. The dynamic information, such as velocity and acceleration, are also successfully extracted. Under experimental conditions, dynamic recognition and tracking were achieved at a speed of 19.55 cm/s and an acceleration of 9.78 cm/s². Additionally, by exploiting the temporal features of static backgrounds and noise, we successfully removed static interference and random noise. The scheme can effectively achieve the separation and imaging of dynamic targets in dynamic scattering media, even when multiple moving targets overlap. This method holds significant potential for dynamic target tracking, recognition, and real-time imaging in complex environments, with broad applicability in fields such as remote sensing and military reconnaissance.

## 2. Methods

## 2.1 Block-based Parallel Image Reconstruction Method Using Compressive Sensing

Single-pixel imaging, based on Compressive Sensing (CS) theory, enables high-dimensional signal reconstruction from minimal measurements, significantly reducing data acquisition requirements and hardware complexity. However, as the demand for higher imaging resolution increases, especially in applications such as satellite remote sensing, astronomical observation, and non-destructive testing, this approach faces limitations in balancing imaging quality with acquisition speed. To address these challenges, BCSI emerges as a significant optimization of SCS, aiming to improve both imaging efficiency and reconstruction accuracy. As shown in Fig. 1(a), this method divides the target's optical signal into several non-overlapping regions (referred to as "blocks"), with independent sampling and reconstruction performed for each block. The optical signal is partitioned using a Spatial Light Modulator (SLM), where the signal of each block is modulated and independently measured by a single-pixel detector. The optical signals within each block are sampled and reconstructed separately, ensuring the independence of each block. The reconstruction process for each block can be expressed by the following equation:

$$y_i = \Phi_i \cdot S_i + n_i \quad (1),$$

where $y_i$ represents the light intensity measurement for the $i$-th block, $\Phi_i$ is the modulation matrix associated with that block, $S_i$ is the sparse representation of the image for that block, and $n_i$ is the noise during the measurement process for that region.

In the image reconstruction process, the results of all block regions must be stitched together according to their positions in the original image. Suppose the image is divided into $N$ sub-regions, and the position of each sub-region in the original image $S_i$ is denoted by $(i_x, i_y)$, where $i_x$ and $i_y$ represent the coordinates of the block region along the horizontal and vertical axes, respectively. The reconstruction process for the final image is expressed as follows:

$$S_{reconstructed} = \sum_{i=1}^{N} S_i(i_x, i_y) \quad (2),$$

where $S_{reconstructed}$ is the final reconstructed image, $S_i(i_x, i_y)$ is the reconstruction result of the *i*-th block region, and each sub-image is placed in the correct position based on the positional information $(i_x, i_y)$ from the original segmentation.

BCSI is essentially a sampling method that performs parallel SCS operations across different blocks. By segmenting the image into independent block regions, the number of measurements is reduced and processed in parallel, significantly enhancing imaging speed while reducing computational complexity. This method overcomes the limitation of prolonged measurement times inherent in traditional imaging techniques, making it particularly suited for fast dynamic scene imaging. For instance, with the 4×4 block modulation scheme, the imaging frame rate increases by a factor of 16 compared to single-pixel compressive sensing, while also substantially reducing the computational complexity of image reconstruction. In fast dynamic scenes, although the parallel sampling and reconstruction strategy significantly improves imaging speed, the quality of the reconstructed image still requires further enhancement. The reasons for this can be attributed to three main factors: (1) SCS is still applied within each block, and the undersampling strategy during acquisition leads to inherent information loss; (2) In low-light conditions, SCS relies on single-photon detectors to achieve high detection sensitivity, but this introduces random noise, such as shot noise and backscatter noise; (3) The limitations of low-pixel detector fabrication result in low pixel utilization between blocks, leading to information loss during block processing, causing edge information loss and blocky artifacts.

Among the challenges mentioned, the blocky artifacts highlighted in the third point are particularly evident during the image reconstruction and stitching process, leading to a noticeable degradation in image quality. Therefore, when employing block parallel methods to enhance imaging frame rates, the effective recovery of missing signals and the elimination of blocky artifacts become critical tasks for achieving high-quality, high-dynamic imaging.

**2.2 Spatiotemporal Deep Learning with U-Net-LSTM Network**

This paper proposes a high-quality block compressed sensing image reconstruction method based on spatiotemporal deep learning. This approach significantly improves the quality of the reconstructed images by effectively integrating spatiotemporal data. The specific method is illustrated in Fig. 1(b). By analyzing the spatiotemporal characteristics of the images, this method accurately exploits inter-frame differences to compensate for signal loss caused by block segmentation and undersampling. Integrating deep learning techniques, we further enhance reconstruction performance through neural network training. Moreover, this approach increases the robustness of image reconstruction by mitigating environmental noise interference, leveraging inconsistencies in spatiotemporal information. For dynamic noise, the random nature of environmental and shot noise allows effective identification and removal. For stationary noise, its static characteristics can be detected through inter-frame consistency, enabling effective suppression.

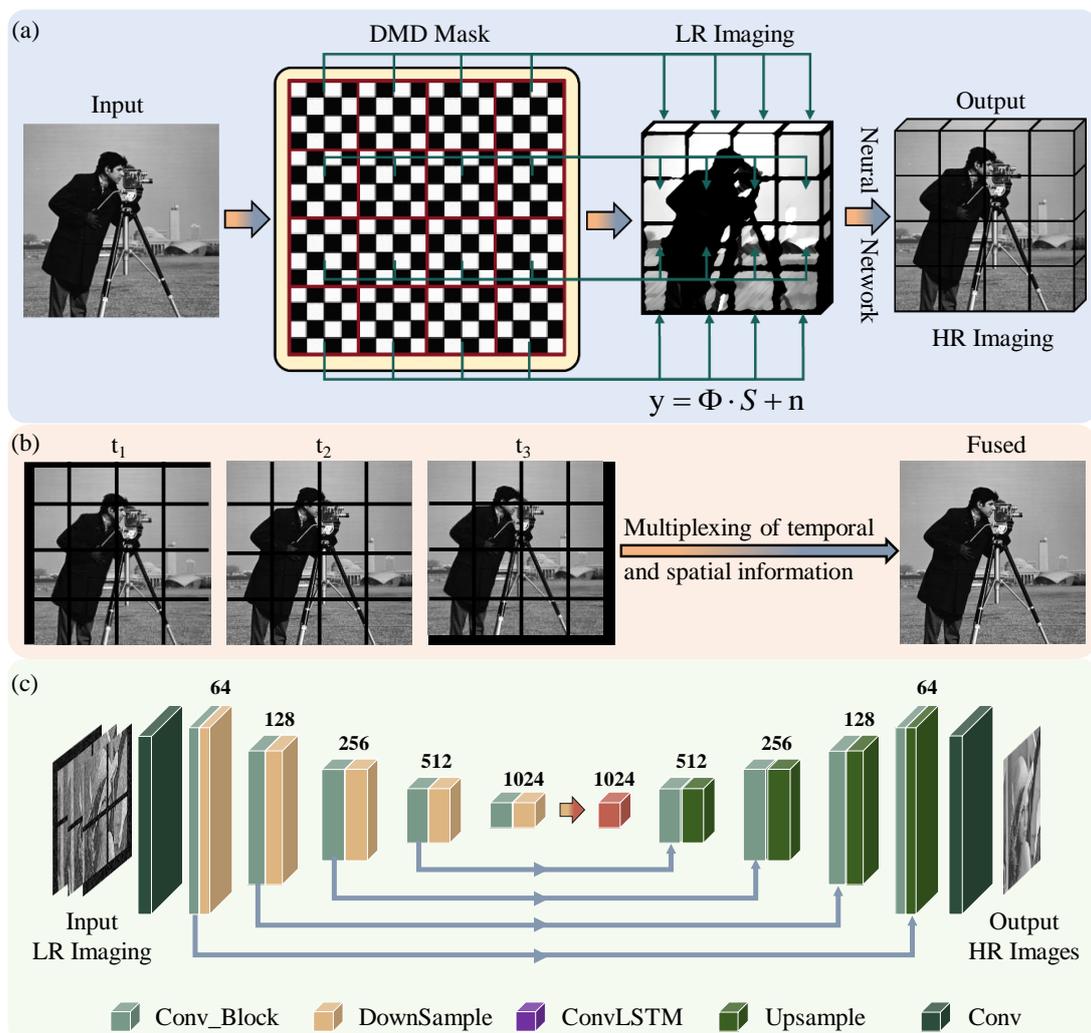

**Fig. 1 Block Compressive Sensing Imaging (BCSI) Principles.** (a) BCSI Principle. (b) Block-wise Dynamic Complementary Principle. (c) U-Net-LSTM Network Architecture.

As shown in Fig. 1(c), we propose a U-Net-LSTM network architecture for image restoration, which combines the structural advantages of U-Net with the temporal modeling capabilities of ConvLSTM. During training, camera-captured images serve as the ground truth, while frames reconstructed using block compressed sensing are provided as input. In this process, the U-Net network excels in global image recovery and fine detail restoration, significantly enhancing the overall image quality. The ConvLSTM network effectively extracts inter-frame temporal information, accurately identifying and removing inconsistencies and noise. In dynamic target imaging, it plays a critical role in suppressing random noise and static interference. Throughout the reconstruction process, the collaboration between U-Net and ConvLSTM facilitates the effective restoration of information from adjacent frames, improving image quality, reducing noise, and recovering more fine details.

In the implementation of the neural network, we employ a comprehensive loss function combining mean squared error (MSE), PSNR, and SSIM for image reconstruction. Specifically, the loss function consists of two components: classification loss and image reconstruction loss. The classification loss is evaluated using the cross-entropy function, which assesses the model's classification ability in the target detection task. The image reconstruction loss is defined as a weighted combination of SSIM, PSNR, and MSE, aiming to optimize the structural similarity of the image, noise suppression, and pixel-level accuracy.

The classification loss is given by:

$$Loss_{class} = \mathrm{CrossEntropy}(c, c_t) \quad (3),$$

where $c$ represents the network's classification prediction, and $c_t$ is the true label in the classification task (for single-target, $c_t = 0$; for dual-target, $c_t = 1$).

The image reconstruction loss is given by:

$$Loss_{recon} = \frac{1}{N} \sum_{i=1}^{N} \left( \alpha \cdot Loss_{MSE}\left(o_1^i, t_1^i\right) + \beta \cdot Loss_{PSNR}\left(o_1^i, t_1^i\right) + \gamma \cdot Loss_{SSIM}\left(o_1^i, t_1^i\right) \right)$$
$$+ c_t \cdot \frac{1}{N} \sum_{i=1}^{N} \left( \alpha \cdot Loss_{MSE}\left(o_2^i, t_2^i\right) + \beta \cdot Loss_{PSNR}\left(o_2^i, t_2^i\right) + \gamma \cdot Loss_{SSIM}\left(o_2^i, t_2^i\right) \right) \quad (4),$$

where $N$ is the batch size, $o_1^i$ and $o_2^i$ represents the output image of the network, $t_1^i$ and $t_2^i$ is the target image, and α, β, γ are the weight coefficients for the MSE,

PSNR, and SSIM loss functions, respectively, with $Loss_{MSE}$, $Loss_{PSNR}$ and $Loss_{SSIM}$ being the corresponding loss functions.

The total loss function is given by:

$$Loss_{total} = \lambda_{recon} \cdot Loss_{recon} + \lambda_{class} \cdot Loss_{class} \tag{5},$$

where $\lambda_{recon}$ and $\lambda_{class}$ represent the relative weights of the reconstruction loss and classification loss in the total loss. During the training process, we used the following hyperparameters: Epoch = 200, batch size = 16, and α, β, γ are set to 0.9, 0.067, and 0.033, respectively, while $\lambda_{recon}$ and $\lambda_{class}$ are 1 and 0.15, respectively. For the learning rate, we adopted a dynamic adjustment strategy, where the learning rate decreases gradually as training progresses, which helps to stabilize the model by the end of training. The initial learning rate was set to 0.0001, and if the validation loss does not improve for 5 consecutive epochs, the learning rate is halved. This adjustment continues until the minimum learning rate reaches 0.0000001. The U-Net-LSTM model was implemented using PyTorch (2.0.0) and trained on a machine equipped with 128 GB RAM, two Intel Xeon Gold C6226R CPUs, and an RTX A6000 GPU.

## 3. Number of Blocks and Imaging Quality

### 3.1 Number of Blocks

Under conditions of low photon counts, single-pixel compressive sensing imaging often faces challenges in balancing reconstruction quality with acquisition speed. The block-parallel strategy reduces sampling time by segmenting the image into smaller, independent processing units. Additionally, by leveraging the spatial sparsity and local structural information of the image, this approach effectively mitigates the information loss caused by insufficient global sampling. Compared to single-pixel reconstruction, the block strategy reduces the number of modulation masks required. This reduction significantly enhances detail recovery and mitigates the impact of noise, ultimately optimizing image quality while maintaining the same sampling conditions. However, while increasing the number of blocks can improve reconstruction quality to some extent, it also introduces potential drawbacks. Firstly, since each sub-block requires independent reconstruction, this increases computational load and complexity. Secondly, excessively fine block divisions lead to insufficient information within each block, hindering effective reconstruction and degrading recovery quality. Moreover, over-partitioning weakens the global data correlation, making it harder to integrate the image's global structure, which further degrades reconstruction quality. Additionally, as the number of blocks increases, memory and storage consumption grow exponentially, placing significant pressure on large-scale image processing. Therefore, optimizing the block strategy to achieve an optimal balance between reconstruction quality and processing time is critical.

In this section, we propose a simulation framework based on BCSI to compare the impact of different block sizes on image reconstruction quality, while keeping the compression rate and sampling time fixed. It is important to note that this framework exclusively employs block-based compressive sensing imaging without incorporating the U-Net-LSTM network architecture discussed earlier. The framework primarily simulates the following key processes: block partitioning, signal modulation, illumination sources, and light signal detection. To implement the simulation, the target image, consisting of $N \times N$ pixels, is first divided into $n \times n$ blocks, with each block containing $\frac{N}{n} \times \frac{N}{n}$ pixels (denoted by matrix elements $(i, j)$). The light signal $\boldsymbol{B}_{\text{blk.}}$ for each block region can be expressed as:

$$\boldsymbol{B}_{\text{blk.}} = \boldsymbol{B} \cdot \boldsymbol{M} \qquad (6),$$

where $\boldsymbol{B}$ represents the intensity of the original image in the block region, and $\boldsymbol{M}$ denotes the modulation mask for the sub-region. In the simulation process, a random sparse matrix is used as the modulation mask for each sub-region, and the complete modulation mask is formed by combining the $n \times n$ sub-region masks.

Assuming the illumination source follows a Poisson distribution, the total photon count rate $\text{K}_{\text{blk.}}$ within a single sub-region can be expressed as:

$$\text{K}_{\text{blk.}} \sim \text{Possion}\left( \sum_{i,j=1}^{N/n} (\boldsymbol{B}_{\text{blk.}}(i,j) \cdot \eta_{\text{dev.}}) \right) \qquad (7),$$

where $\boldsymbol{B}_{\text{blk.}}(i,j)$ represents the intensity of the modulated $(i,j)$-th pixel, and $\eta_{\text{dev.}}$ denotes the device's response error rate, which accounts for photon counting errors caused by the superposition of pulse responses when multiple photon signals arrive within a short time. The photon count $\text{P}_{\text{det.}}$ collected by a single detection unit over a time interval $t$ can be expressed as:

$$\text{P}_{\text{det.}} = \sum_{0}^{t} \text{K}_{\text{blk.}} + \text{P}_{\text{noise}} \qquad (8),$$

where $P_{noise}$ represents the noise counting rate during the detection process, which includes readout noise, dark current noise, background noise, and shot noise. In the simulation, readout noise follows a normal distribution, dark current noise follows a Poisson distribution, and both background and shot noise are modeled with a normal distribution. From this, the imaging data of the target image after the physical acquisition process is obtained. Subsequently, the TVAL3 algorithm is used for image reconstruction, restoring the block detection and imaging process.

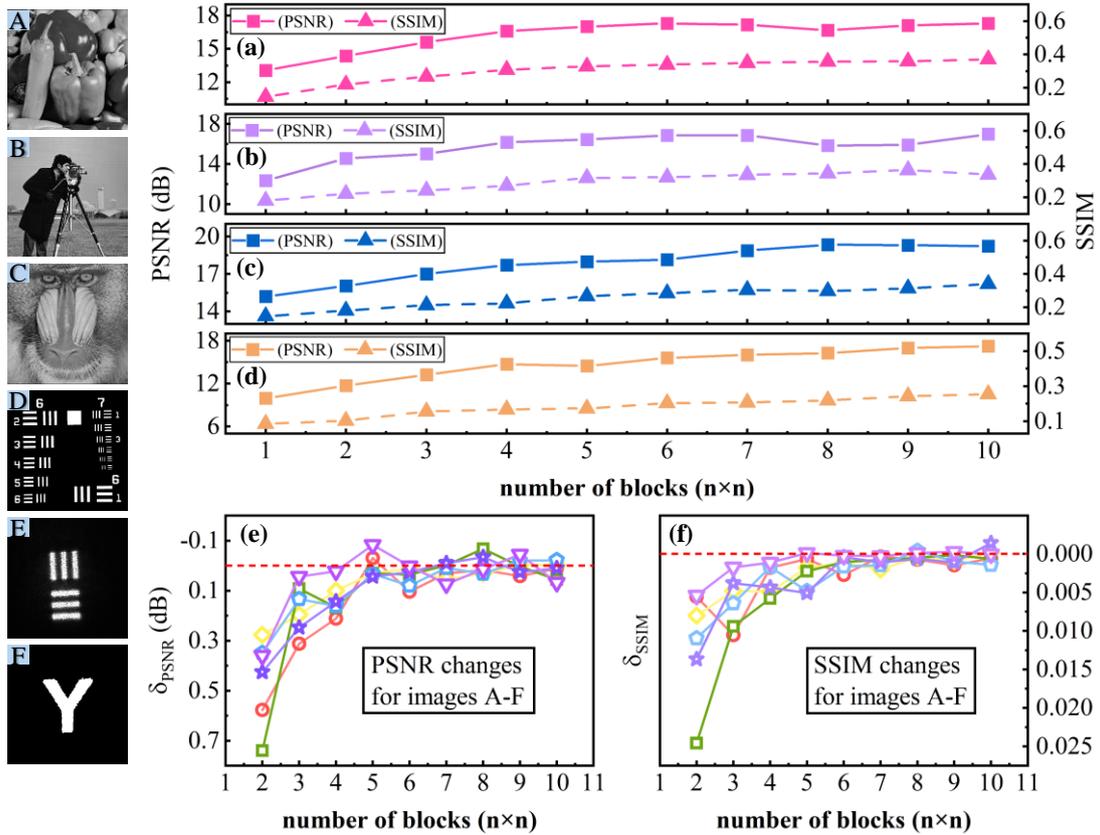

**Fig. 2 Image Reconstruction Effects with Different Numbers of Blocks.** (a-d) Block Reconstruction Quality of Four Different Targets (A-D). (e, f) The Effect of Number on PSNR and SSIM. (A-F) Seven Classic Images Used in the Simulation.

As shown in Fig. 2, we simulate the image reconstruction effects for different block numbers under a photon-level illumination environment. In Fig. 2(a-d), a compression rate of 25% is applied using a random sparse matrix to modulate and reconstruct the target images A-D on the left. The results indicate that, as the number of blocks increases, both the PSNR and SSIM of the target images improve, demonstrating that, for the same sampling time, BCSI delivers higher image quality than traditional SCS. However, as the number of blocks continues to increase, the improvements in PSNR and SSIM gradually plateau.

To clarify this result, we applied the same operation to the seven images (A-F) shown in Fig. 2 and analyzed how PSNR and SSIM vary with the number of blocks for each image. The changes in image PSNR ($\delta_{PSNR}$) and SSIM ($\delta_{SSIM}$) caused by each additional block are used as indicators to assess the positive impact of increasing block numbers on image quality. The corresponding results are shown in Fig. 2(e) and (f). It can be observed that transitioning from single-pixel imaging to 2×2 block partitioning significantly improves image quality. However, starting from a 4×4 block partition, the improvement in image quality due to increased block numbers gradually diminishes, indicating that further partitioning may not lead to substantial improvements. Additionally, we examined the effect of block numbers on image quality under varying illumination intensities. The results show that the improvement in image quality with increased block numbers is more pronounced at lower illumination intensities, due to the reduced signal-to-noise ratio in dim lighting conditions. Block modulation allows for a longer modulation time per mask during the same acquisition period, which enhances the signal-to-noise ratio. However, when the illumination intensity is sufficiently high, block modulation no longer results in significant improvements to the signal-to-noise ratio. For more details, please refer to Fig. S2.

**3.2 Optical Experiment**

Based on the previous discussion, we conclude that 4×4 block partitioning provides the optimal balance between imaging quality and speed under the experimental conditions of this study. Therefore, we selected a 4×4 block photon-level detector (MPPC, Hamamatsu, C13369-3050EA-04) for the subsequent experiments. The experimental setup is shown in Fig. S3. The sample is moved using an electrically controlled two-dimensional translation stage. After the sample is illuminated by a laser, the scattered light is reflected by a fast steering mirror (FSM, Newport, FSM-300 Series) and directed to the Digital Micromirror Device (DMD, UPOLabs, HDSLM136D70-DDR, 1024×768). The DMD loads masks to spatially modulate the optical signal, with a resolution of 256×256. Signals are collected using a charge-coupled device (CCD, MVCAM, AI-030U780M, 640×480) and an MPPC, with the CCD capturing the true image of the object as the Ground Truth. The signals from the MPPC are recorded by a time-correlated single-photon counting system (TCSPC, Siminics, MT1616) and transmitted to a computer for further processing.

First, we conducted imaging experiments on targets with small amplitude movements. The targets were a resolution chart and the letter "Y". Fig. 3(a) and (e) show the standard images of the two targets captured by the CCD. At a sampling rate of 6.25%, we performed high-speed imaging at $256\times256$ pixels and 50 fps for two different dynamic targets. Fig. 3(b) and (f) show the images of the targets directly reconstructed using the block compressive sensing algorithm, where blocky artifacts due to partitioning are clearly visible. Compared to the standard images, the PSNR values are 7.11 dB and 6.14 dB, and the SSIM values are 0.1725 and 0.0395, respectively. While the contours of the targets are still recognizable, the images suffer from significant noise, blocky artifacts, and inter-block seams. These imaging quality issues are primarily caused by the noise effects of the photon-level detector in parallel single-pixel imaging. Subsequently, based on the spatiotemporal deep learning scheme we proposed, we exploit the data differences between consecutive frames to recover the fine details and overall structure of the targets. The high-quality reconstruction, further refined by the U-Net-LSTM network, is shown in Fig. 3(c) and (g). Additionally, in Fig. 3(d) and (h), we analyze the lateral resolution of the imaging results. The recovery performance of our method is nearly identical to that of the standard images, validating the effectiveness of our approach. Compared to traditional BCSI, the spatiotemporal deep learning block compressive sensing method, when combined with deep learning, significantly improves image quality. The PSNR improves by 28.5 dB, and the SSIM increases by 0.86, providing strong evidence of the advantages of this approach in high-speed multi-pixel imaging. Moreover, high-quality images at high frame rates allow us to capture clear dynamic information. As shown in Fig. 3(i) and (j), images of the resolution chart and the letter "Y" are presented at 0.1s, 0.2s, 0.3s, 0.4s, 0.6s, 0.8s, and 1.0s. The upward and downward movements of the resolution chart, along with the random motion of the letter "Y," are clearly visible.

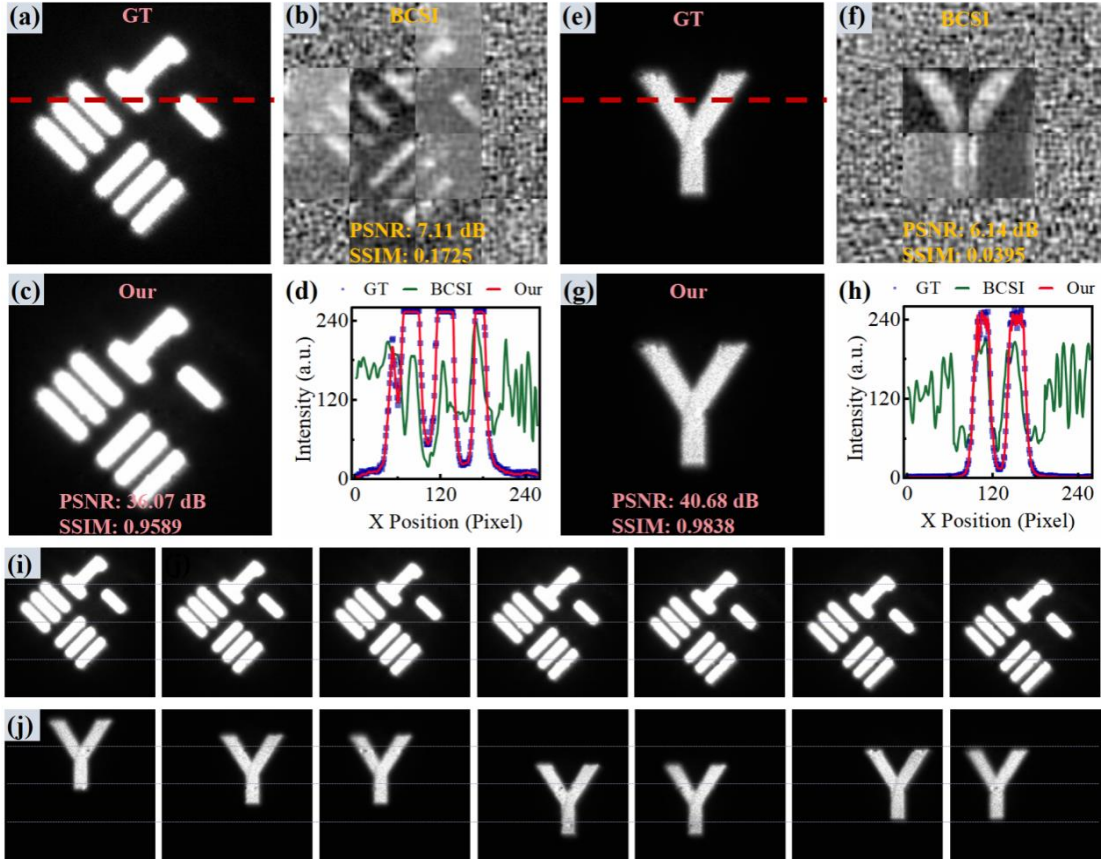

**Fig. 3 Dynamic Target Imaging Recovery Results.** (a), (e) Standard images of the targets. (b), (f) Recovery results using BCSI. (c), (g) Recovery results using U-Net-LSTM. (d), (h) Lateral intensity distributions at the red dashed line positions in (a) and (e) for GT, BCSI, and our method. (i), (j) Recovery results of the targets at 0.1 s, 0.2 s, 0.3 s, 0.4 s, 0.6 s, 0.8 s, and 1.0 s.

By further adjusting the compression ratio to 3.125% and the modulation frequency to 25.6 kHz, we achieved dynamic imaging at 200 fps. Meanwhile, by applying our U-Net-LSTM network to recover the BCSI imaging results, high-quality, high-dynamic imaging was obtained. This demonstrates the feasibility of capturing object motion information and analyzing dynamic parameters using this approach. To this end, we performed imaging and tracking of objects with complex motion. Fig. 4(a-d) shows the direct BCSI results and those obtained using our method for the petal-shaped motion of the car model and the spiral motion of the airplane model. The motion trajectories extracted from the dynamic video are shown in Fig. 4(e) and (f). Subsequently, based on our approach, we extracted the dynamic information of both moving targets. As shown in Fig. 4(g), within the imaging field of view, the car and airplane models move along different trajectories simultaneously. We tracked their position changes over time in the two-dimensional space, as shown in Fig. 4(h). The resulting velocity and acceleration trajectories, presented in Fig. 4(i), align with the experimentally preset motion parameters. Under the experimental conditions, the dynamic target with a speed of 19.55 cm/s and an acceleration of 9.78 cm/s² can be identified. This indicates that the proposed method provides rich information for analyzing the target's motion.

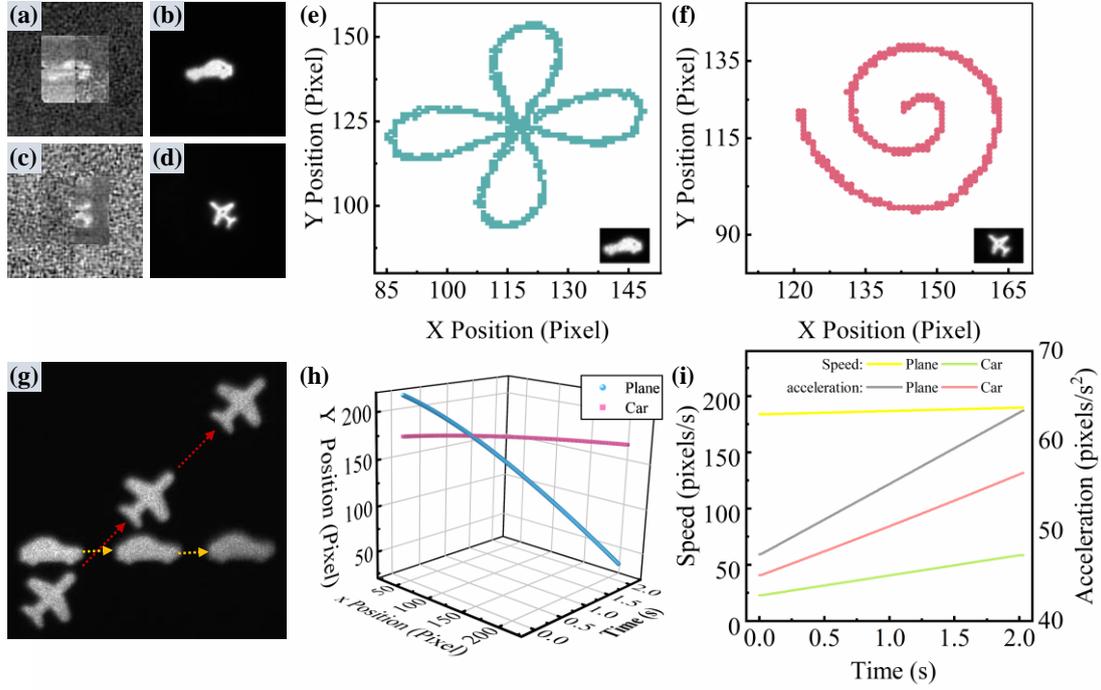

**Fig. 4 Dynamic Target Trajectory Recognition.** (a), (c) BCSI imaging results of the targets (car, airplane). (b), (d) Target images reconstructed using U-Net-LSTM. (e), (f) Constant-speed motion trajectories of dynamic targets. (g) Schematic of the dual dynamic target movement. (h) Motion trajectories of the dual dynamic targets over time. (i) Velocity and acceleration of the dynamic target during the acceleration process as functions of time.

Additionally, in real dynamic tracking scenarios, the presence of static backgrounds and dynamic noise significantly affects imaging quality. The key feature of this approach is its ability to exploit the differences between consecutive frames, along with the structural similarity of the target, to achieve high-quality imaging. Since the position of the static background remains constant during dynamic imaging, static interference can be effectively suppressed. This is accomplished by modeling the motion trajectory of the dynamic target during inter-frame information restoration using the U-Net-LSTM network. Environmental and device noise, which are typically irregular, can be efficiently removed by leveraging spatiotemporal deep learning during dynamic imaging.

To validate this hypothesis, we first conducted dynamic imaging experiments in a laboratory with a complex static background. The imaging targets were two cars moving toward each other, set against a background of buildings, a lighthouse, streetlights, and trees. This scene was simulated by controlling the reflection angles of the DMD micromirror. Figs. 5(a-b) show the imaging results with and without a static background, respectively. From top to bottom, the images include the standard CCD images, directly reconstructed BCSI images, and the images of the two cars recovered using our proposed method. It can be observed that our approach effectively removes the static background. Even when the spatial overlap between the two cars exceeds 90%, dynamic information enables the clear extraction of dual target images, providing an effective tool for tracking and monitoring. Subsequently, we introduced dynamic water mist in front of the imaging system using a mist generation device to simulate a dynamic thin scattering imaging scenario. Fig. 5(c) and (d) display the imaging results with and without a static background under the influence of the water mist. The imaging conditions at the same position, excluding the water mist, are identical to those in Fig. 5(a). As expected, both the CCD-captured ground truth (GT) and BCSI images show significant deterioration under the influence of the water mist. However, the method proposed in this study still significantly enhances the image reconstruction of dynamic targets, clearly capturing the target's positional information. This result demonstrates that, despite the challenges posed by complex backgrounds, our method can optimize dynamic target reconstruction by effectively leveraging spatiotemporal information fusion. It validates the robustness and efficiency of our approach in complex dynamic scenarios.

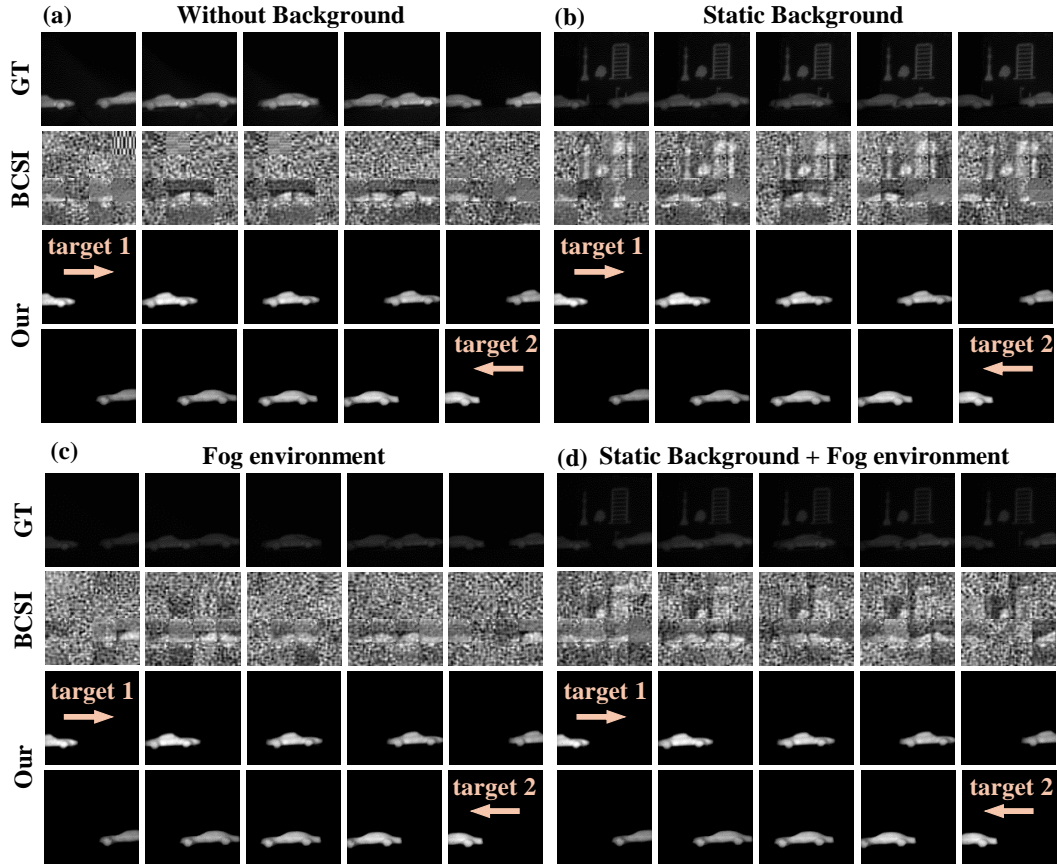

**Fig. 5 Imaging Experiment of Dual Dynamic Targets in a Complex Environment.** (a-d) Show the recovered images of the two dynamic targets moving towards each other under four different scenarios: no background, complex static background, no background with a thin scattering medium, and complex static background with a thin scattering medium. From top to bottom, the images represent the results of GT, BCSI, and our method under different overlap conditions. From left to right, the images depict the various overlap situations during the motion, from the cars moving towards each other to their overlap and finally to moving in opposite directions.

## 4. Conclusion

This paper introduces a high-quality BCSI scheme based on spatiotemporal deep learning. The method fully exploits inter-frame differences to mitigate the information loss in BCSI. Furthermore, we propose the U-Net-LSTM network to address artifacts, and image degradation in conventional BCSI under high-speed dynamic scenarios. Compared to SCS, the proposed approach achieves a 16-fold increase in frame rate under the same sampling rate and modulation frequency, with a maximum imaging speed of 200 fps, significantly enhancing both image detail and clarity. Compared to traditional BCSI, the proposed method improves the PSNR and SSIM by 28.5 dB and 0.86, respectively, demonstrating a substantial advantage in image restoration and quality enhancement. The experimental results demonstrate that the reconstruction method, leveraging spatiotemporal deep learning exhibits excellent robustness and high efficiency when confronted with complex scenes. It effectively recovers image details, suppress noise and reduces artifacts even in the presence of static interference, dynamic thin scattering medium or multi-target overlapping scenarios. Additionally, spatiotemporal deep learning enables the effective extraction of dynamic information, such as the speed and acceleration of moving targets. This scheme provides more reliable technical support for target recognition and trajectory tracking of fast-moving targets. By further optimizing the scheme to improve imaging resolution and real-time processing capability, it is anticipated that the application will be extended to more complex environments.

**Competing interests:** The authors declare that they have no competing interests.

**Acknowledgements:** The authors gratefully acknowledge support from the National Natural Science Foundation of China (Nos. U23A20380, U22A2091, 62222509, 62005150, 62127817, and 6191101445), National Key Research and Development Program of China (Grant No. 2022YFA1404201), Shanxi Province Science and Technology Innovation Talent Team (No. 202204051001014), 111 projects (Grant No. D18001) and Shanxi Provincial Basic Research Program Project (202203021222107).